\documentclass[aip,apl,reprint]{revtex4-1}

\usepackage{chemformula} 
\usepackage[T1]{fontenc} 
\usepackage{graphicx}
\usepackage{textcomp}
\usepackage{fourier}
\usepackage{verbatim}
\usepackage[utf8]{inputenc}
\usepackage[english]{babel}
\usepackage{soul}
\usepackage{xcolor}

\begin{document}

\title{Magnetic Properties of Epitaxially-grown SrRuO\textsubscript{3} Nanodots}

\author{G. Laskin}
\thanks{These two authors contributed equally}
\author{H. Wang}
\thanks{These two authors contributed equally}
\author{H. Boschker}
\author{W. Braun}
\author{V. Srot}
\author{P. A. van Aken}
\author{J. Mannhart}
\email[corresponding author, ]{office-mannhart@fkf.mpg.de}
\affiliation{Max Planck Institute for Solid State Research, Heisenbergstr.\ 1, 70569 Stuttgart, Germany}


\begin{abstract}
We present the fabrication and exploration of arrays of nanodots of SrRuO\textsubscript{3} with dot sizes between 500\,nm and 15\,nm.
Down to the smallest dot size explored, the samples were found to be magnetic with a maximum of the Curie temperature \textit{T}\textsubscript{C} achieved by dots of 30\,nm diameter.
This peak in \textit{T}\textsubscript{C} is associated with a dot-size-induced relief of the epitaxial strain, as evidenced by scanning transmission electron microscopy.
\end{abstract}

\maketitle

\section{Introduction}
Nanostructures of complex functional materials provide an experimental platform to investigate novel physical phenomena if the characteristic device size reaches intrinsic length scales, such as the inelastic scattering length or the Fermi wavelength.
For mean-field systems, especially semiconductors, confinement effects have been studied for decades and have already been implemented in widespread technological applications \cite{Ashoori_QD,Kastner_QD}.
Recently, a significant number of investigations have shown that such confinement of correlated systems in 2D \cite{Bibes_thin_films}, \textit{e.g.}, at interfaces \cite{Mannhart_interfaces_electronics, Hwang_interfaces, Baiutti_2d_supcond}, in thin films \cite{Yoshimatsu_QW_SVO} or superlattices \cite{Bern_SRO_properties}, promises a variety of interesting advantages over their mean-field counterparts.
Although novel electronic properties are expected \cite{Mannhart_dots} if correlated systems are confined in fewer dimensions, \textit{i.e.}, to 1D and 0D, such systems have been explored only marginally.

This work focuses on the limiting case of spatial confinement into 0D objects, \textit{i.e.}, into quantum dots or artificial atoms.
In Refs.~\citenum{Cen_AFM_writing} and~\citenum{Levy_SET} it was shown that such dots can be fabricated at the LaAlO\textsubscript{3}/SrTiO\textsubscript{3} interface using surface-modification with an atomic force microscope (AFM) tip.
Ion-beam based fabrication of magnetic SrRuO\textsubscript{3} nanodots with diameters of about 80\,nm has been reported in Ref.~\citenum{Ruzmetov_SRO_dots}.
In Ref.~\citenum{Klein_SRO_islands} the magnetization reversal in similar nanostructures of various sizes was studied by magnetic force microscopy. 
We here use the top-down approach based on epitaxial growth and electron-beam lithography that allows such objects to be fabricated with high precision and reproducibility.
At the same time, the form and mutual arrangement of these objects may be chosen and varied as desired.
The main challenge for fabricating such devices is to synthesize a highly pure material with low defect density that along with small sizes allows the formation of a coherent many-body electron wave function.
However, it is well known that the patterning process causes local material damage and dead layers, which can significantly reduce the mean free path of the electrons \cite{Katine_dead_layers}.

In order to study the material behavior upon confinement in nanodots, we chose SrRuO\textsubscript{3} \cite{Koster_review} as a model system. 
SrRuO\textsubscript{3} is a correlated conducting perovskite oxide with strong itinerant ferromagnetism (bulk Curie temperature \textit{T}\textsubscript{C}  $\approx 160$\,K) and an intensively studied perovskite material.
It exhibits orthorhombic symmetry at room temperature.
The ferromagnetic properties of SrRuO\textsubscript{3} depend highly on the material quality as well as on the film thickness \cite{Ishigami_SRO,Chang_SRO_thick_limit}.
Even slight variations of the stoichiometry or lattice parameters may lead to a significant change of magnetic properties \cite{Siemons_SRO_stoich, Lu_strain_SRO, Thomas_SROonDSO}.
Hence, SrRuO\textsubscript{3} is a model system to study the relevance of dead-layer creation and how materials behave upon confinement within low-dimensional objects.
We therefore prepared epitaxial SrRuO\textsubscript{3} nanodots with various diameters and explored how the magnetic properties of the material depend on the size of the structures.

\section{Experimental Techniques}

For the sample preparation, special care was taken to deposit SrRuO\textsubscript{3} films of high purity and with the desired stoichiometry.
Prior to SrRuO\textsubscript{3} deposition, SrTiO\textsubscript{3} substrates were terminated \textit{in situ} in the growth chamber by thermal annealing at 1300\,\textdegree C for 200\,s at a molecular oxygen pressure of 0.08\,mbar \cite{Jaeger_APL}.
Immediately after that, and in the same UHV chamber, SrRuO\textsubscript{3} thin films were grown by pulsed laser deposition with a target-substrate distance of 56 mm.
The substrate temperature during growth was 680\,\textdegree C, and the molecular oxygen pressure in the chamber was kept at 0.08\,mbar.
The SrRuO\textsubscript{3} target was ablated using a KrF excimer laser with a wavelength of 248\,nm and an energy density on the target of 2.5\,J/cm\textsuperscript{2}.
The nanodots were patterned using a commercial 100\,keV electron beam lithography system (JEOL JBX6300) and CSAR (Allresist, AR-P 6200) as resist material.
After resist development, a 20-nm layer of amorphous Al\textsubscript{2}O\textsubscript{3} was deposited, which was then used as a hard mask for dry etching with Ar ions at an energy of 600\,eV and a pressure of $1.4\times10^{-4}$ mbar. 
For a pattern design, we used rectangular arrays of dots, where the dot periodicity equalled two dot diameters. Arrays covered the entire substrate area, excluding edge regions.
For 20-nm dots, $\approx 10^9$ dots were fabricated on each substrate.

Scanning transmission electron microscopy (STEM) specimens in cross-sectional orientation were prepared by focused ion beam (Zeiss CrossBeam XB 1540) cutting, then nanomilled at liquid nitrogen temperature (Fischione NanoMill Model 1040).
STEM investigations were carried out using a spherical aberration-corrected microscope (JEOL JEM-ARM 200F) with a DCOR probe corrector (CEOS GmbH) at 200 kV.

\section{Results}

\begin{figure}
\includegraphics[width=0.9\columnwidth]{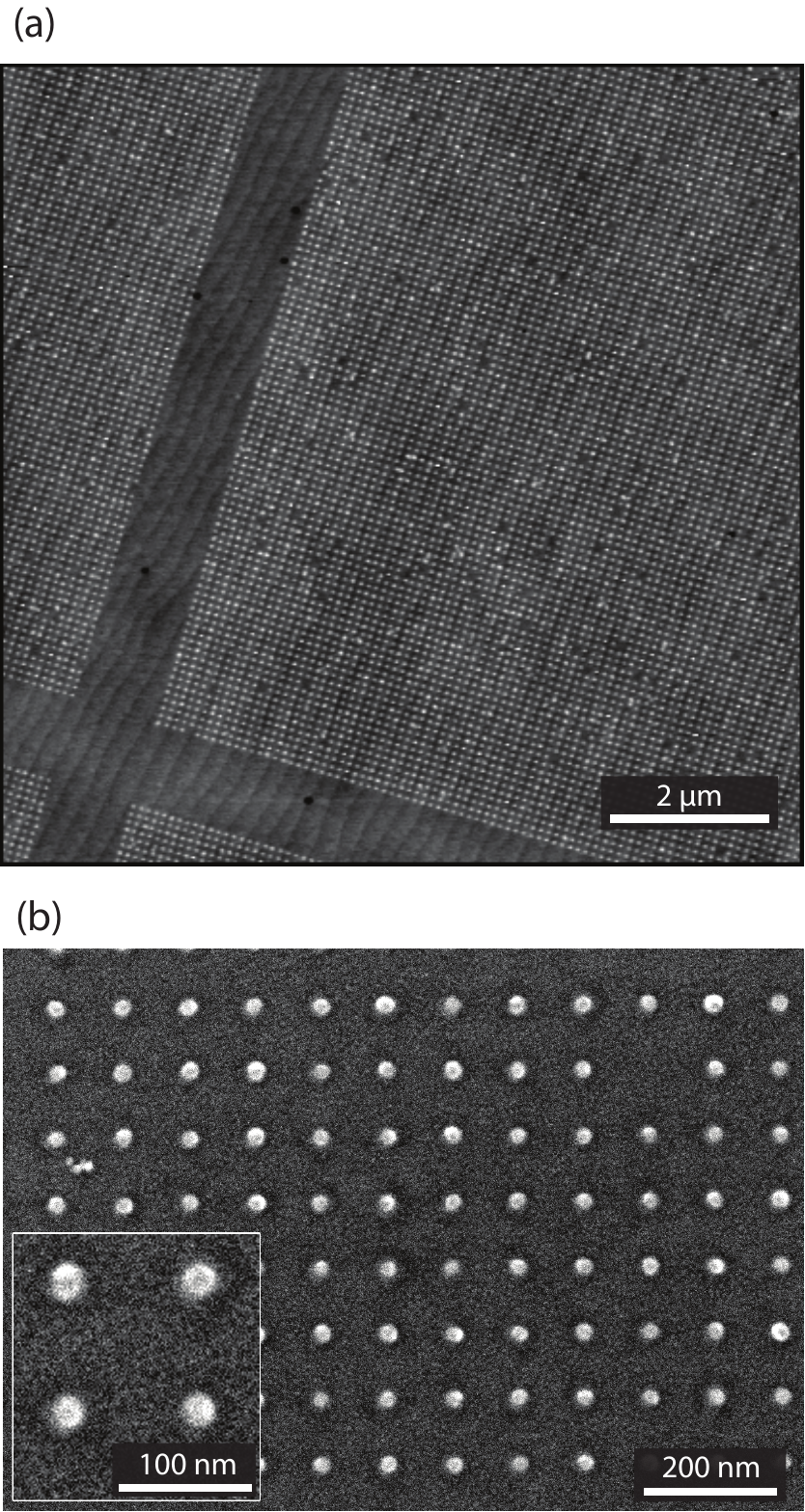}
\caption{(a) AFM image of SrRuO\textsubscript{3} dot arrays grown on SrTiO\textsubscript{3} showing both the 1-unit-cell high substrate surface steps and the dot arrays.
The height of the dots is approximately 3\,nm.
(b) Top view SEM image of a SrRuO\textsubscript{3} dot array with a diameter of 20\,nm and height of 12\,nm. \label{fig:SEM}}
\end{figure}

\begin{figure}
\includegraphics[width=0.9\columnwidth]{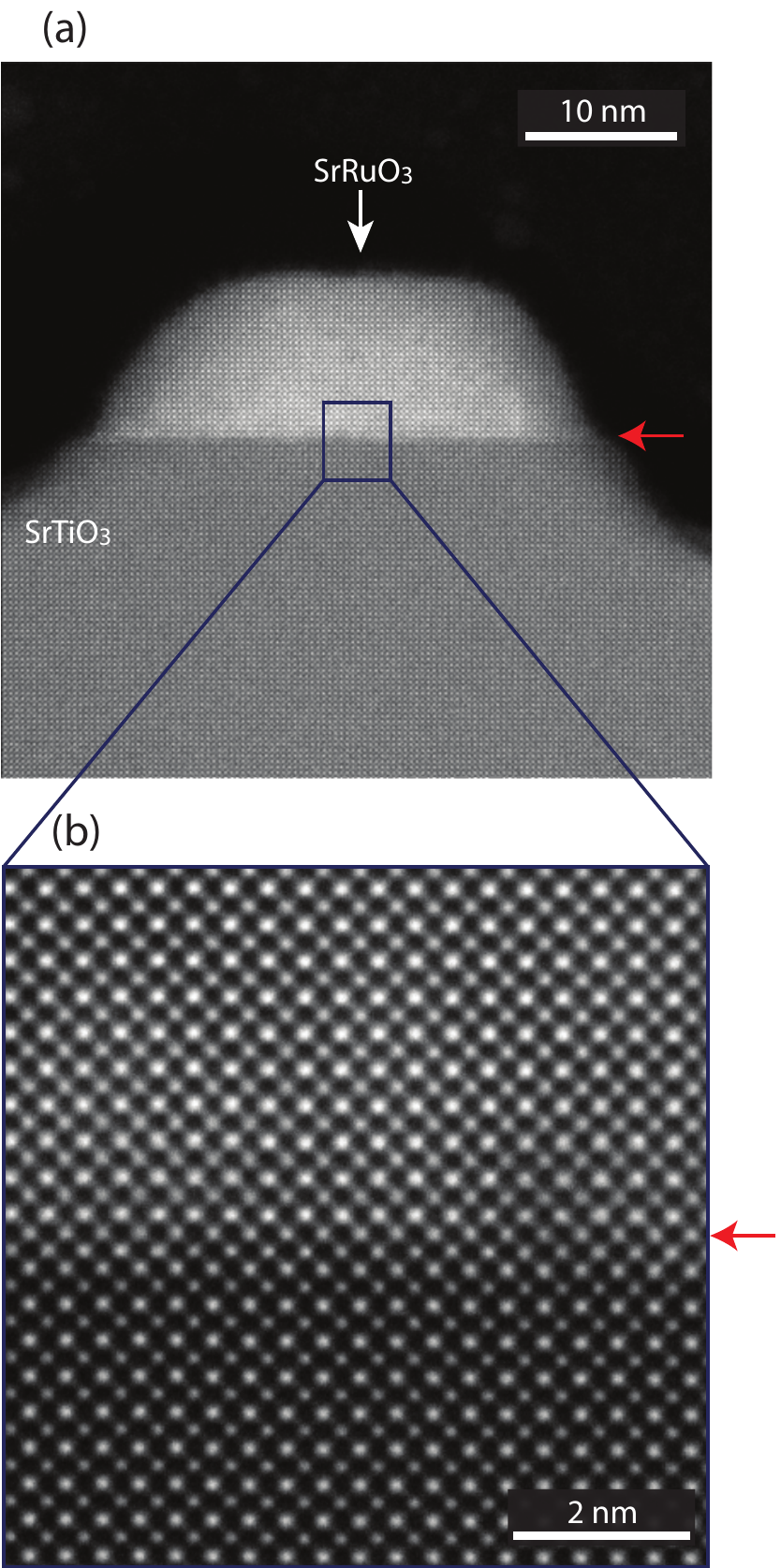}
\caption{(a) \textit{Z}-contrast STEM image of a cross-sectional cut through a single dot close to its center.
(b) Magnified STEM image showing the interface (red arrow) between the dot and the substrate. \label{fig:StEM}}
\end{figure}

Figures \ref{fig:SEM} and \ref{fig:StEM} give an overview of the dots' microscopic properties.
Figure \ref{fig:SEM}(b) is a scanning electron microscopy (SEM) image of a sample with a rectangular array of 20-nm SrRuO\textsubscript{3} dots.
A \textit{Z}-contrast STEM image of a single-dot cross section from a different sample is shown in Fig.~\ref{fig:StEM}(a), which depicts a clear contrast between the SrRuO\textsubscript{3} and the substrate material with a sharp interface. 
This is magnified in Fig.~\ref{fig:StEM}(b).
During the sample preparation for TEM, the dot was cut through its center with a lamella thickness of $\approx 20$\,nm.
Therefore, the cross section shown in Fig. \ref{fig:StEM}(a) provides a projection through almost the entire dot.
The curved shape of the substrate near the dot and the shape of the dot itself are caused by non-uniform removal of material during dry etching with Ar ions. 

The magnetic properties of different samples were measured using a SQUID magnetometer (Quantum Design MPMS) equipped with the reciprocating sample option (RSO) head. 
For all samples, the thickness of the SrRuO\textsubscript{3} film equaled 12\,nm (\textit{i.e.}, 32 unit cells of SrRuO\textsubscript{3}).
Every sample was measured twice at different steps of the fabrication process: directly after the thin-film growth and then after being patterned into nanodots. 
To measure the zero-field-cooled and the field-cooled behavior of the magnetization, the samples were cooled to 4\,K in zero magnetic field, after which an out-of-plane magnetic field of 0.1\,T was applied. 
The samples were then warmed to room temperature and cooled again.

\begin{figure}
\includegraphics[width=0.9\columnwidth]{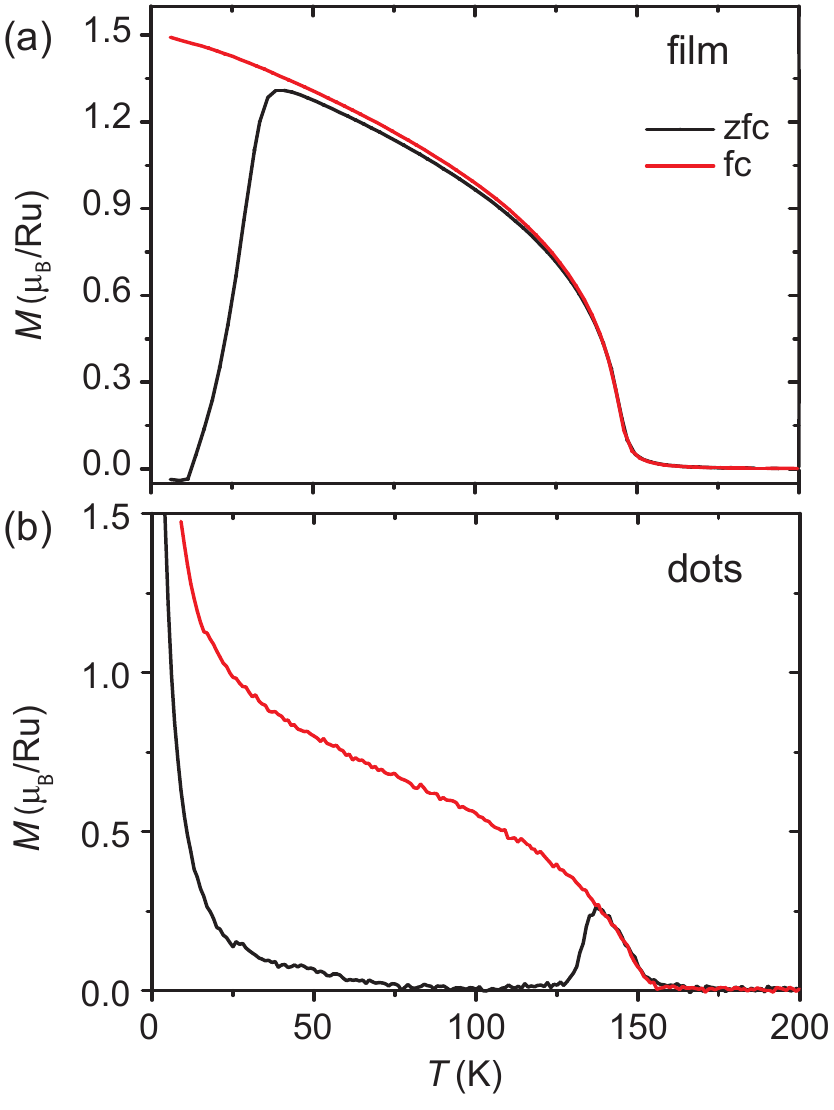}
\caption{Temperature-dependent magnetization of (a) an unpatterned 12-nm-thick SrRuO\textsubscript{3} film and (b) an array of 20-nm-diameter dots patterned from the same film, measured with zero-field cooling (zfc) and field-cooling (fc) in an out-of-plane oriented magnetic field of 0.1\,T.
The black and red lines correspond to zero-field-cooled and field-cooled measurements, respectively.
The diamagnetic contribution of the SrTiO\textsubscript{3} substrate has been subtracted. \label{fig:M(T)}}
\end{figure}

A typical characteristic of the temperature-dependent magnetization of a SrRuO\textsubscript{3} film is shown in Fig.~\ref{fig:M(T)}(a).
The shape of the curve and the saturation magnetization of 1.5 \textit{$\mu$}\textsubscript{B}/Ru are in good agreement with the literature data \cite{Tian_SRO_properties,Grutter_SRO_properties}.
The Curie temperature \textit{T}\textsubscript{C} is determined by the linear extrapolation of the transition curve to $M=0$, although a finite magnetic moment exists above this temperature, attributed to the domain structure and film reorientation \cite{Xia_SRO}.
For all films, the transition has been found at a \textit{T}\textsubscript{C} = 149$\pm$1\,K, which is comparable to the \textit{T}\textsubscript{C} values reported in the literature for equally thick films deposited by pulsed laser deposition \cite{Bern_SRO_properties,Chang_SRO_thick_limit}.
Fig. \ref{fig:M(T)}(b) shows an example of the magnetic transition curve for a patterned sample with a dot diameter of 20\,nm.
Considerably different is the behavior of the magnetic response at low temperatures, where a significant increase of the magnetization is observed.
This is due to the paramagnetism inherent in SrTiO\textsubscript{3} \cite{Coey_STO_mag}, which causes difficulties for measuring the dot magnetization because of the very small SrRuO\textsubscript{3} volume compared to the SrTiO\textsubscript{3} substrate.
However, this contributes only to the low-temperature properties and does not affect the magnetization close to the ferromagnetic transition region, in which we are primarily interested here.

\begin{figure}
\includegraphics[width=0.9\columnwidth]{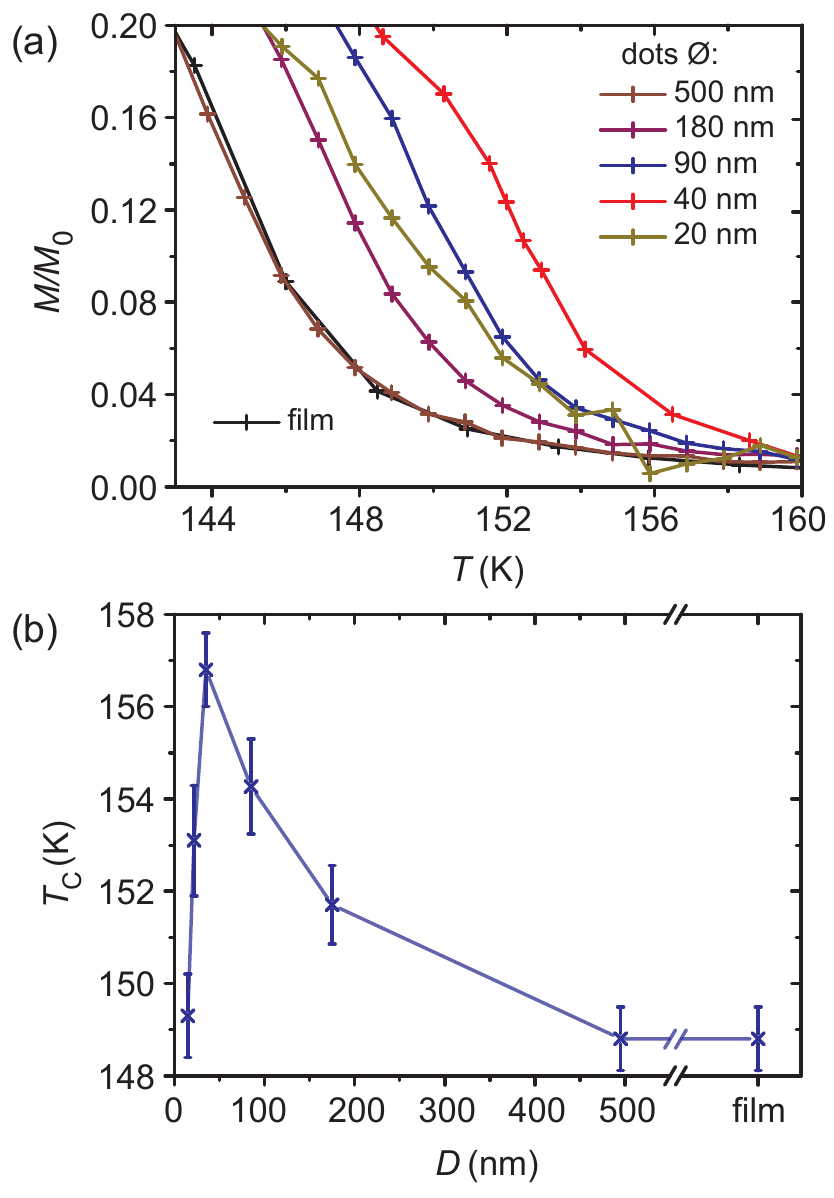}
\caption{(a) Measured temperature dependence of the magnetization of several samples with different dot sizes in the temperature range close to \textit{T}\textsubscript{C}.
All curves are normalized by the magnetization of the film.
(b) Curie temperature of the samples plotted as a function of dot diameter.\label{fig:M(T)_all}}
\end{figure}

\begin{figure*}
\includegraphics[width=2.0\columnwidth]{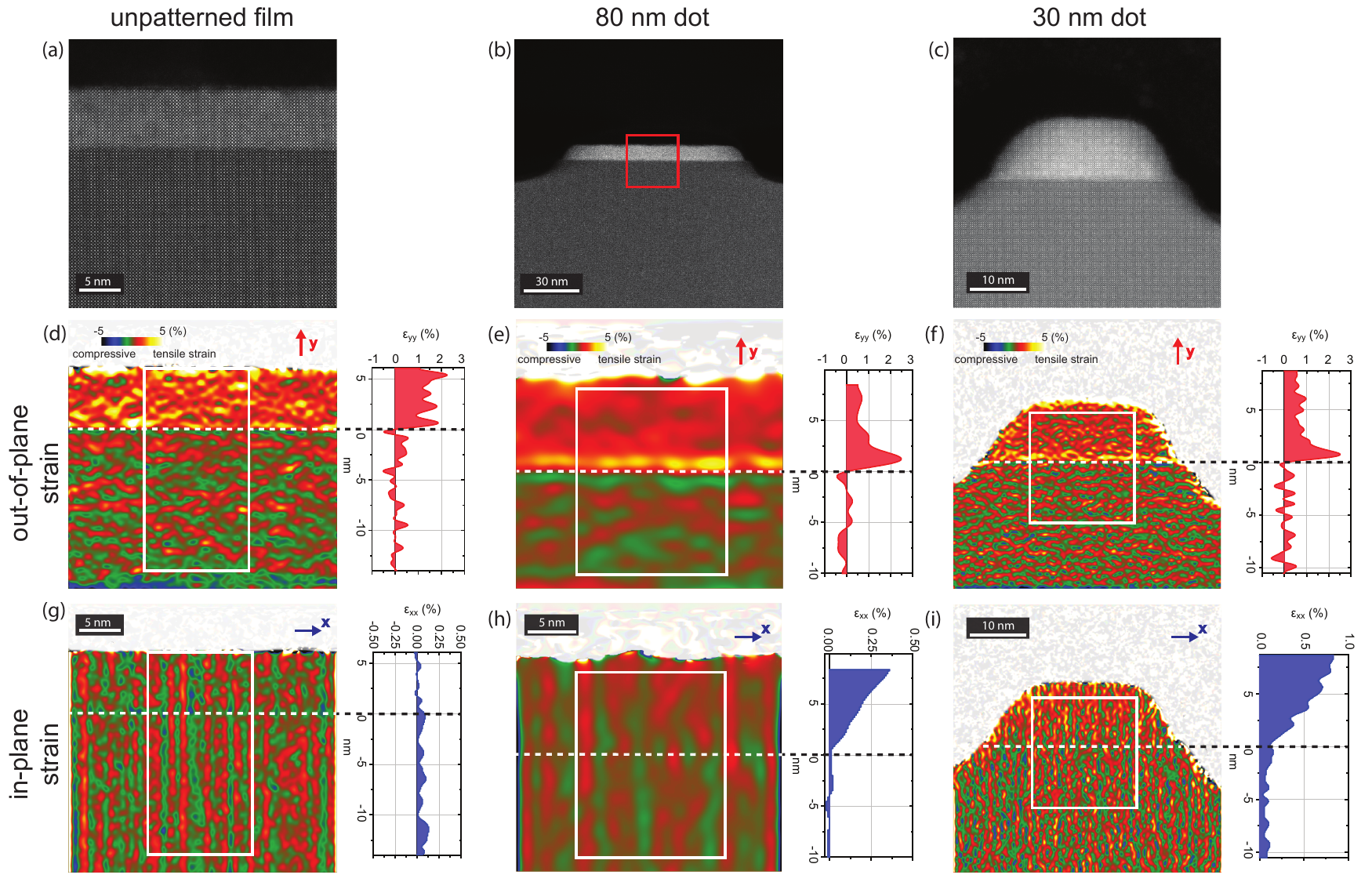}
\caption{Original \textit{Z}-contrast STEM images are shown for (a) a SrRuO\textsubscript{3} unpatterned thin film, (b) a 80-nm-sized dot and (c) a 30-nm-sized dot on (100) SrTiO\textsubscript{3}.
The respective out-of-plane $\varepsilon\textsubscript{yy}$ (d,e,f) and in-plane $\varepsilon\textsubscript{xx}$ (g,h,i) components of the strain are obtained from STEM-based geometric phase analysis.
For the 80-nm-sized dot, the central region of the dot (red rectangle) is analyzed.
The data in the graphs is extracted from the horizontally averaged regions marked by the boxes.
The 0\% level corresponds to the lattice constant of bulk SrTiO\textsubscript{3}. \label{fig:strain}}
\end{figure*}

After patterning, the sample remains ferromagnetic with a \textit{T}\textsubscript{C} = 153\,K, \textit{i.e.}, 4\,K higher than that of the film.
This increase of the Curie temperature is systematically observed for dots of different sizes.
To clarify this effect in more detail, a series of samples with different dot sizes (from 500 to 15\,nm) was fabricated.
The initial film morphology and their magnetic properties were similar.
With decreasing dot size, the distance between them was scaled proportionally to keep the total amount of SrRuO\textsubscript{3} constant.
After patterning, the temperature dependence of the magnetization was measured for all samples using the procedure described above.
The results of the measurements for dots of different sizes are depicted in Fig.~\ref{fig:M(T)_all}(a) and the extracted variation of the Curie temperature in Fig.~\ref{fig:M(T)_all}(b).
The \textit{T}\textsubscript{C} does not change after patterning for dots equal to or larger than 500\,nm.
For smaller dots, a gradual increase of \textit{T}\textsubscript{C} with decreasing dot size is observed.
This is valid down to a dot size of approximately 30\,nm, after which \textit{T}\textsubscript{C} starts to decrease, presumably due to surface defects and dead layers.
A decrease of \textit{T}\textsubscript{C} with shrinking feature size is well known for thin films \cite{Chang_SRO_thick_limit,Himpsel_magnetic_nanostructures_review} and has been observed in nanoparticles of other perovskites \cite{Vasseur_LSMO_nanoparticles} and also for metals \cite{Rong_FePt_nanoparticles}.
This reduction of \textit{T}\textsubscript{C} is generally attributed to the reduced number of neighbors and the corresponding destabilization of the magnetic ordering \cite{Rong_FePt_nanoparticles}.
Even for the smallest dot sizes, however, the Curie temperature exceeds that of the unstructured film.

Strain is one of the mechanisms known to influence magnetic properties of SrRuO\textsubscript{3}.
Using \textit{ab-initio} density functional theory (DFT) calculations, it was shown in Ref.~\citenum{Zayak_SRO_strain_DFT} that uniaxial and epitaxial strain in SrRuO\textsubscript{3} significantly alters its magnetic properties.
Experimental studies of SrRuO\textsubscript{3} grown on different substrates \cite{Lu_strain_SRO} demonstrate that the magnetic properties differ depending on the amount of strain in the film.
Furthermore, SrRuO\textsubscript{3} films released from their substrates demonstrate a 10\,K higher \textit{T}\textsubscript{C} of 160\,K \cite{Gan_SRO_relieved}.
A different study \cite{Thomas_SROonDSO} showed that a SrTiO\textsubscript{3} capping layer grown on SrRuO\textsubscript{3} epitaxial films deposited on DyScO\textsubscript{3} modifies  the oxygen octahedral structure in the SrRuO\textsubscript{3} with a corresponding increase of the Curie temperature.
To explore whether strain relief is responsible for the \textit{T}\textsubscript{C} increase in our dot structures, we studied the strain in the dots by STEM.
We used geometric phase analysis to analyze high spatial resolution STEM images in the in-plane and out-of-plane directions \cite{Hytch_STEM_1,Hytch_STEM_2}. 
The lattice constant of the SrTiO\textsubscript{3} substrate was taken as a reference level.
Fig.~\ref{fig:strain}(d,g) shows the strain distribution of a SrRuO\textsubscript{3} thin film grown on SrTiO\textsubscript{3}.
The data reveals that the film is fully strained, \textit{i.e.}, the lattice constant in the in-plane direction remains unchanged across the interface (Fig.~\ref{fig:strain}(g), right panel), whereas in the out-of-plane direction the unit cell of SrRuO\textsubscript{3} is elongated by a constant value of  $\approx$1.4$\pm$0.3\,\% compared to SrTiO\textsubscript{3} (Fig.~\ref{fig:strain}(d), right panel).
These strain characteristics are altered by patterning the epitaxial film into nanodots, as shown by Fig.~\ref{fig:strain}(e,h) for a 80-nm-diameter dot and Fig.~\ref{fig:strain}(f,i) for a 30-nm-diameter dot.
For both dot sizes, in the in-plane direction (Fig.~\ref{fig:strain}(h,i)), starting from the interface, the lattice constant of SrRuO\textsubscript{3} increases gradually, which indicates strain relaxation.
For the 80-nm dot, the lattice constant has relaxed by 0.35$\pm$0.10\,\% at the top of the dot, whereas for the 30-nm dot the relaxation is even stronger. It reaches 0.75$\pm$0.10\,\%, matching fully relaxed SrRuO\textsubscript{3}.
We link this lattice relaxation to the observed increase of the Curie temperature.
In the out-of-plane direction (Fig.~\ref{fig:strain}(e,f)) the behavior for dots of both sizes is similar. Here, in the first two unit cells close to the interface, we observed the strain close to that of the film, which, as expected, then decays together with the lateral strain relaxation to approach a three-dimensionally relaxed structure.
A similar mechanism of patterning-induced strain relaxation has been observed before in patterned semiconductor nanostructures, for example Si \cite{Himcinschi_Si_strain_relief}, \mbox{InGaN/GaN} \cite{Ramesh_InGaN_strain_relief} or (Ga, Mn)As \cite{Wenisch_GaMnAs_strain_relief}.
We therefore conclude that the strain relaxation is responsible for the increase of the \textit{T}\textsubscript{C} in these SrRuO\textsubscript{3} nanodots.

\section{Summary}
In summary, using e-beam lithography of epitaxial SrRuO\textsubscript{3} films, we have succeeded in fabricating nanosized dots of SrRuO\textsubscript{3} ranging in size from several hundred nm down to 15\,nm.
SrRuO\textsubscript{3} shows ferromagnetism even in the smallest dots. 
Unexpectedly, magnetometry reveals that the magnetic properties are enhanced for dot sizes below 500\,nm, which manifests itself in an increase of the Curie temperature for small dots.
We demonstrate that the Curie temperature depends on the size of the dots, increasing gradually from 149 K in the thin film and dots larger than 500\,nm to 157\,K at 30\,nm dot size.
This behavior is attributed to strain relaxation in the material caused by the removal of lateral constraint around the dot.

\section{Acknowledgements}

GL acknowledges E. Goering (MPI for Intelligent Systems, dept. Modern Magnetic Systems) and E. Bruecher (MPI for Solid State Research, Chemical Service) for the support with magnetometry measurements. We thank T. Reindl, U. Waizmann and J. Weis for the technical support with sample fabrication. Discussions with J. Mydosh and M. Ternes are gratefully appreciated. HW gratefully acknowledges the China Scholarship Council (No. 201504910813) and the Max Planck Society for financial support. We thank Y. Wang, U. Salzberger, K. Hahn and P. Kopold for their support during TEM sample preparation and the TEM experiments. We acknowledge the assistance by B. Fenk for preparing FIB lamellas and by L. Jin from the Ernst Ruska-Centre Juelich for the nanomill preparation of TEM lamellas.

\bibliography{laskin_nanodots}

\end{document}